\documentclass[a4paper]{article}

\usepackage{graphicx}
\usepackage{epstopdf}
\usepackage{abstract}

\title{Towards a More Complete Object-Orientation in Design Grammars}
\date{2020\\ March}
\author{Samuel Vogel \thanks{Hochschule Ravensburg-Weingarten}, Peter Arnold \thanks{IILS mbH and external doctorate University Stuttart}} 

\begin{document}

\maketitle

\begin{abstract}
The ongoing digital transformation in industry applies to all product life cycle's stages. The design decisions and dimensioning carried out in the early conceptual design stages determine a huge part of the product's life cycle costs (LCC). The automation of the conceptual design phase promises therefore huge gains in terms of LCC. Design grammars encode design processes in production systems made up of rule sequences which automatically create an abstract central product model (central data model) from given requirements. Graph-based design languages use the Unified-Modeling-Language (UML) to define the product entities (classes) supporting object-oriented inheritance. Graphical rules instantiate the classes and iteratively assemble the central model. This paper proposes to extend the design languages by introducing methods (operations). This allows the use of object-oriented design patterns and interface mechanisms as object-oriented principles are then fully implemented. A graphical mechanism to model the method calls is presented which integrates seamlessly into the graph-based design language's graphical rule specification. The object oriented design grammar enables modularization and reusability of engineering knowledge. The integration of engineering domains is enhanced and multistakeholder collaboration with access control (information security) becomes feasible.
\end{abstract}

%%%%%%%%%%%%%%%%%%%%%
\section{Introduction}
%\paragraph{Rough Motivation of the stuff}
The digitalization of the early design stages promises a huge efficiency gain in product development~\cite{Pah2007}. A great amount of the product life cycle costs is determined in the early conceptual design phase's decisions. Advanced digital engineering methods provide a digital product mock-up for optimization and validation in the virtual world, before producing the first real-world product embodiment (''do it right the first time''). The interdisciplinary systems engineering approach tries to capture all aspects of a product (design) and supports to handle the complexity in developing modern products~\cite{Wal2015}. Product and software engineering has to grow together to enable the development of mechatronics products and systems.

%%%%%%%%%%%%%%%%%%%%%
\section{State of Knowledge and Previous Work}
%\paragraph{What did you/others do? }
The main idea behind systems engineering is to decompose a complex product, as well as its requirements, into smaller systems that are made up of (sub)systems themselves~\cite{Kom2011}.  Clear interfaces are defined between linked systems, that can be interchanged afterwards according to the interfaces' definitions. The decomposition leads to smaller system entities that can be handled more easily and enables therefore the handling of complex products and systems (divide-and-conquer).\\

Model-based systems engineering using object-oriented modeling languages as \emph{UML}~\cite{Uml2014} or \emph{SysML}~\cite{SML2017} is a heavily used tool in systems engineering that naturally supports the system-of-systems decomposition through its object-oriented features~\cite{Boo2007}:
\begin{itemize}
 \item \emph{Abstraction:} Representing only the essential features of an entity. Extraction of the essential features and interfaces of a system.
 \item \emph{Encapsulation:} Wrapping entities that are defined by data and methods to conduct specific behavior into units. Hiding system internals and implementations behind externally accessible and well defined interfaces. 
 \item \emph{Polymorphism:} Sub-typing of entities through hierarchical inheritance. Behavior is abstractly defined by interfaces that are implemented and reused in sub-types. 
\end{itemize}
Reusable design patterns are widely spread in software engineering. These patterns heavily rely on the object-oriented features listed above and follow the central idea of: \emph{``Programming to an Interface, not an Implementation''}~\cite{Gam1994}. The interface mechanism is the central feature of flexible object-oriented software design which allows to easily couple, exchange and reuse ``black-box'' entities whose interaction is specified through interfaces and which is independent of the inner structure and specific implementation of the entities. This fits perfectly to the systems engineering central idea to recursively couple encapsulated and hierarchically defined sub-systems to compose increasingly complex systems and products.\\

State-of-the-art systems engineering work flows rely mainly on a manual creation of a central product model in a modeling language aforementioned. The level of automation in modern engineering tools is usually limited to the application of macro and API (application programming interfaces) functionalities for sub tasks. In advanced system engineering approaches model-to-model transformations are introduced. Simulation and validation models are automatically derived from a manually created central data model~\cite{Rei2016}. Other methods have been examined in science to increase the level of autonomy and automation, especially in the early product design phase. Knowledge-based engineering methods have been used to realize a fully autonomous concept design~\cite{Lar2012,Sob2015}. Formal (design) grammars are established to automate design processes and to implement even complex design tasks and make them (re-)executable~\cite{Ant2001}.\\

\begin{figure}[!ht]
\centering
\includegraphics[width=0.8\textwidth]{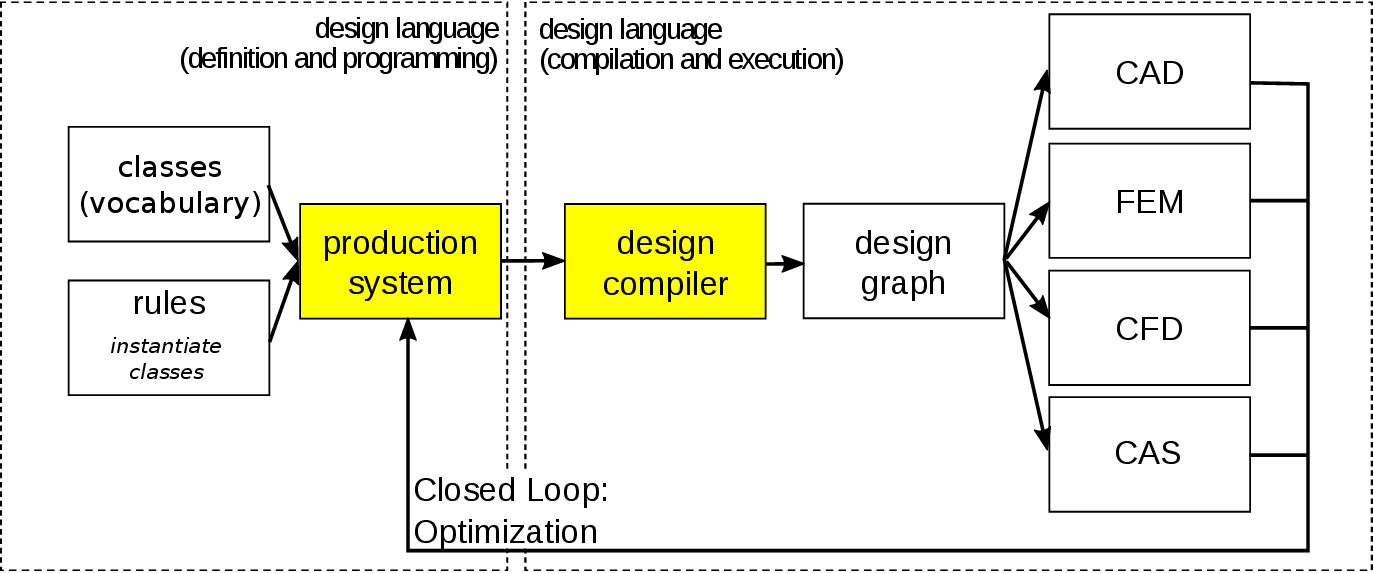}
\caption{Information architecture of a design grammar (design language).}
\label{fig:infoarchitectur}
\end{figure}
Graph-based design grammars are a generic approach that combines both, object-oriented modeling of the product entities in a UML class diagrams~\cite{Her2007} and a production system made up of a rule sequence~\cite{Kroe2005}. Muenzer presents a different approach of a graph-based and object-oriented grammar to model product design processes~\cite{Mue2015}. Product variants are generated by combinatorial (constraint-based and boolean satisfiability) variation in contrast to the rule based decision logics of graph-based design grammars, that allow a generic modeling of design loops and inner iterations in the design process. A specific product instance is iteratively expanded from a given set of requirements. This is practically done by the execution of the production system in a so-called design compiler~\cite{DC43} that creates a design graph as digital twin of the product. Figure~\ref{fig:infoarchitectur} shows the information architecture of graph-based design grammars schematically. The left part contains the implementation of the design process in the design compiler. Product building blocks are defined in terms of \emph{classes} (historically called \emph{vocabulary}). The \emph{production system} is made up of an adaptive rule sequence that instantiates the classes. The sequence can be branched, based on judgments rendered in \emph{decision nodes}. \emph{Rules} are be defined graphically or as program code, which are able to conduct manipulations of the \emph{design graph}. The design graph acts as central data model (single source of truth). On the right side the validation part of the design process (process chains) is shown where domain-specific engineering models (CAD, structural mechanics, fluid mechanics, etc.) are derived from the central model by executing model-to-model (M2M) transformations. These transformations are implemented in plug-ins that are provided by the design compiler. The design graph is analyzed in the M2M-transformations within the plug-in calls, the domain-specific information is filtered and the domain- and application-specific models are generated and executed. The results of the automatically executed and post processed validation calculations and simulations are fed back into the production system for realizing inner design iterations and outer optimization loops, for example to explore design spaces. Design grammars have been used throughout different applications like automotive, aerospace and manufacturing~\cite{Arn2012, Haq2004,Ira2005,Sch2005,Vog2012,Gro2012,Ton2017}.\\

%%%%%%%%%%%%%%%%%%%%%
\section{Problem Setting}
\label{sec:problemSetting}
\begin{figure}[htp]
\centering
\includegraphics[width=1.0\textwidth]{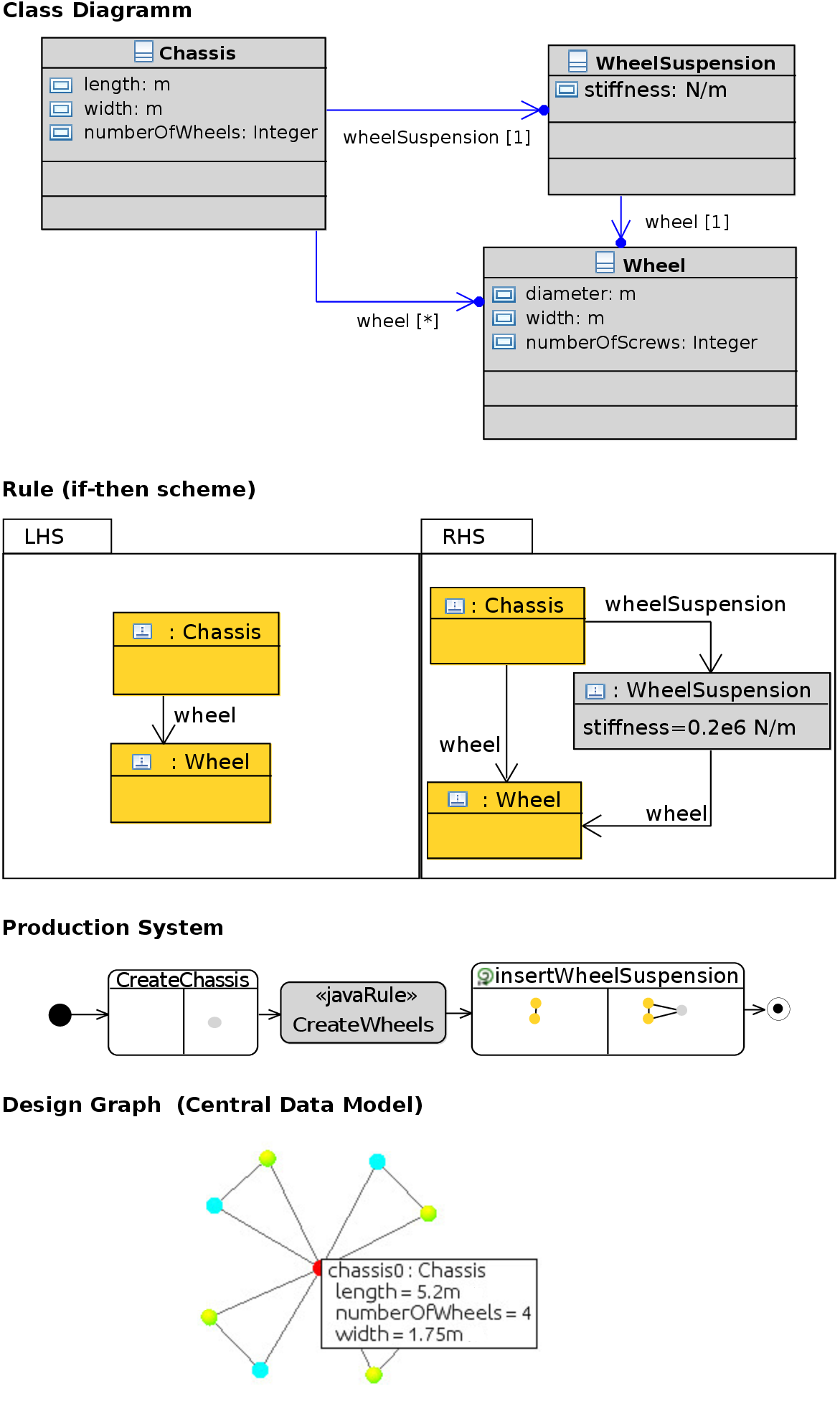}
\caption{Schematic design grammar of a simplified car design.}
\label{fig:desLang}
\end{figure}

Modern systems engineering requires a high degree of collaboration and modularity to cope with multi-physical, interdisciplinary and highly coupled complex systems. The design grammars presented above and shown in figure~\ref{fig:infoarchitectur} are modeled in an object-oriented modeling language (UML), but follow more or less a procedural programming paradigm~\cite{Whi2005}. This is shown in the schematic design grammar illustrated in figure~\ref{fig:desLang}.
A simple car model, consisting of a chassis with a defined number of wheels, is expanded and refined by introducing wheel suspensions. The objects that are put together by the rules are defined in the \textit{UML class diagram} on the top. A graphical rule (if/then scheme, left-hand-side [LHS]/right-hand-side [RHS] scheme) to add \emph{wheel suspensions} between the \emph{wheels} and the \emph{chassis} is shown below. The production system on the lower part shows the rule sequence for iteratively building up the car model. The \emph{JavaRule} hosts a program code rule which iteratively adds the wheels in a loop. It is called \emph{numberOfWheels} times as set in the \emph{Chassis} object. The design graph (central data model) on the bottom of figure~\ref{fig:desLang} is generated by the execution of the production system. It hosts the instances of the current product design with its specific design parameters.\\

\begin{figure}[!ht]
\centering
\includegraphics[width=0.8\textwidth]{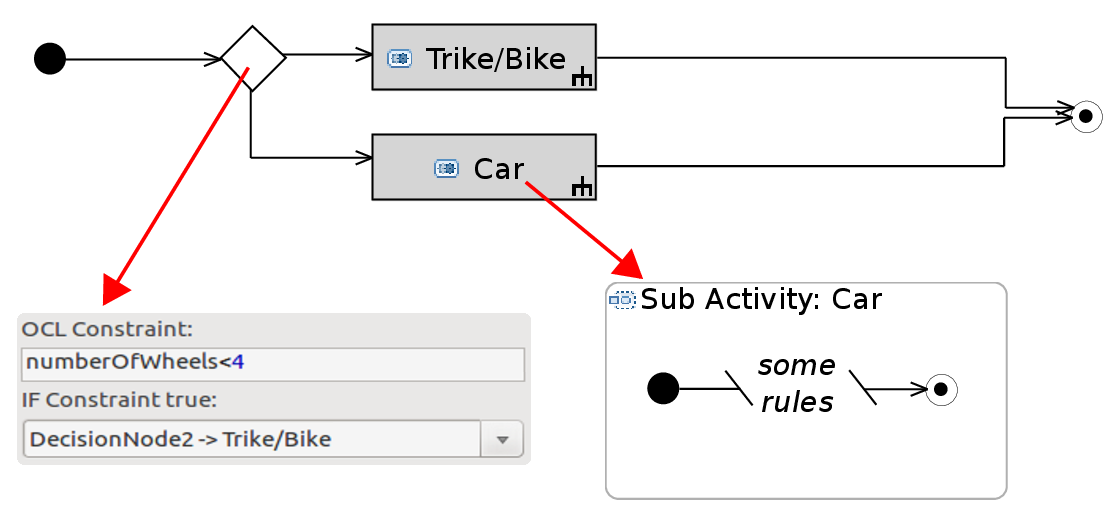}
\caption{\emph{Decision node} for branching the rule sequence (left) based on model constraints and sub activity (right) to model hierarchically embedded rule sequences.}
\label{fig:activiyDiagramm}
\end{figure}
The rule sequence in the production system manipulates the design graph in a procedural manner as imperative commands work on a common program state in a predefined sequence. The production systems in design grammars have an additional entity called \emph{Decision Node} that allows a branching of the rule sequence and the implementation of conditional, switch or loop statements in the design grammar (figure~\ref{fig:activiyDiagramm}). Hierarchical \emph{sub activities} can be modeled in the production system. They can embed sub production systems (figure~\ref{fig:activiyDiagramm}). A sub activity can be seen as being equivalent to a sub routine that takes the central data model as sole argument. In the world of the object-oriented software engineering this approach would be considered as ''bad design''. Translated to object-oriented software engineering, the design grammar's rule sequence can be interpreted as a sequence of static methods - without any explicit method parameters - that builds up the central data model. The lack of both, explicit interface definitions and methods that are coupled to data objects (classes), harms reusability and modularization. For bigger and complex design grammars it gets difficult to maintain consistency and to debug the model as the whole design graph is exposed to every rule and sub activity. Object-oriented modeling realizes the paradigms presented above. An abstract interface mechanism addresses the issues of modularization, reusability and maintainability. \emph{So the lack of a tight encapsulation in conjunction with the missing mechanism of abstractly defined interfaces can be seen as a central challenge of the traditional design grammar approach.} This disadvantage also applies to numerous other expert systems, as rule-based or logical systems, that miss hierarchical modeling concepts~\cite{Jac1998}. The following issues can arise from the described lack of object-orientation:
\begin{itemize}
  \item \emph{Modularization:} Systems-of-systems aspect difficult to realize without explicit interface definitions as the sub systems are difficult to delimit mutually. The behavior of sub entities is detached from the themselves as operations are not coupled to the data entity they apply to. 
  \item \emph{Reusability:} Components can not be properly encapsulated into entities with explicit and self-explanatory interface definitions for later reuse.
  \item \emph{Domain Integration:} The integration of domain-specific models via M2M-transformations is not sufficient. Some domains can be easier integrated when granular class methods (with defined explicit parameter lists bounded to a class) are provided in the production system and the domain-specific models themselves are created iteratively.
  \item \emph{Collaboration:} Designing complex designs needs involvements from multiple domains and therefore involvement of multiple experts. Proper interfaces to clarify the requirements and responsibilities are a prerequisite for successful collaboration.
  \item \emph{Information Security:} Modules with defined interfaces can be encrypted and hidden to allow collaboration between companies whilst respecting intellectual property. 
\end{itemize}
%
%%%%%%%%%%%%%%%%%%%%%
\section{Methods}
\label{sec:methods}
%\paragraph{How did you do it?}
In this work modifications to the design grammar are suggested to increase the level of object-orientation. The proposed solutions have been prototypically implemented on the basis of the Design Compiler 43V2.

\subsection{Code Integration}
\label{subSec:codeIntegration}
\begin{figure}[!ht]
\centering
\includegraphics[width=0.8\textwidth]{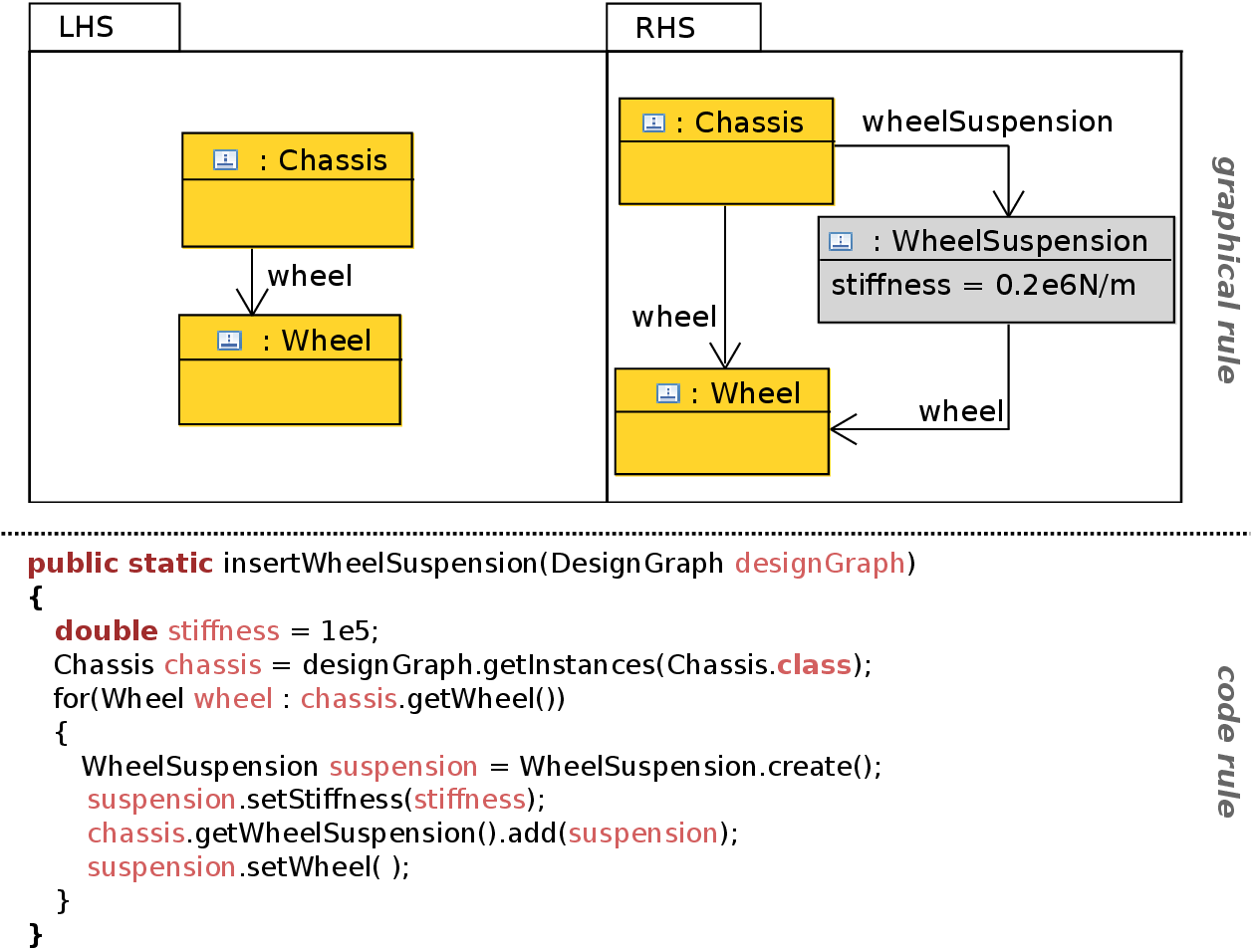}
\caption{Rule \emph{insertWheelSuspension}: Equivalence of graphical rule (top) and code rule (bottom). Both implementation techniques execute equivalent manipulations on the design graph.}
\label{fig:codeTwin}
\end{figure}

The foundation of the following approaches is a tight integration of graphical modeling and code implementation. Graphical manipulation rules of the design graph, as shown in figure~\ref{fig:codeTwin} top, can be equivalently implemented as Java code, figure~\ref{fig:codeTwin} bottom. For each UML class a Java class source file is internally created which hosts the class parameters as well as the setter and getter methods. This enables the user to model the design process either graphically or code-based, depending on the personal preferences or on the specific problem. The graphical and code-based approaches are reduced to just two different views on the same modeling task.

\subsection{Class Methods}
\begin{figure}[!ht]
\centering
\includegraphics[width=0.8\textwidth]{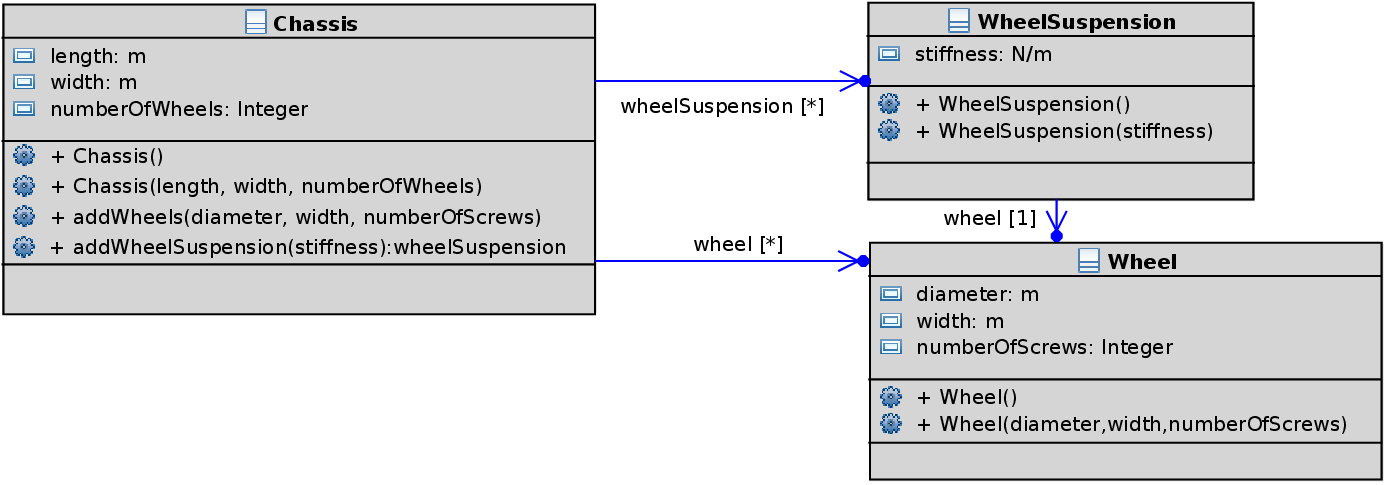}
\caption{Extended class diagram with object-oriented class methods and class constructors (compare with 'classical' class diagram in figure~\ref{fig:desLang}). The data types of the methods' parameters are not shown for the sake of simplicity. This applies to all figures.}
\label{fig:classWithMethods}
\end{figure}

As shown in figure~\ref{fig:classWithMethods}, methods are added to the classes in the class diagram to enable \textit{encapsulation}. The introduction of classes' methods realizes the object-oriented foundation principle of a tight coupling between data and methods. Constructors of the classes can be also defined. These are executed in the creation of the classes' instances. The methods are executed on selected instances of the class in the production system. Both, constructors and methods can be called from source code or graphically, see subsection~\ref{subSec:callMethodsAndConstructors}. Based on subsection~\ref{subSec:codeIntegration}, the constructors and methods can be implemented either as source code or graphically as shown in subsection~\ref{subSec:modelingMethods}.

\subsection{Modeling Abstract Interfaces}
\label{subSec:interfaces}
\begin{figure}[!ht]
\centering
\includegraphics[width=0.8\textwidth]{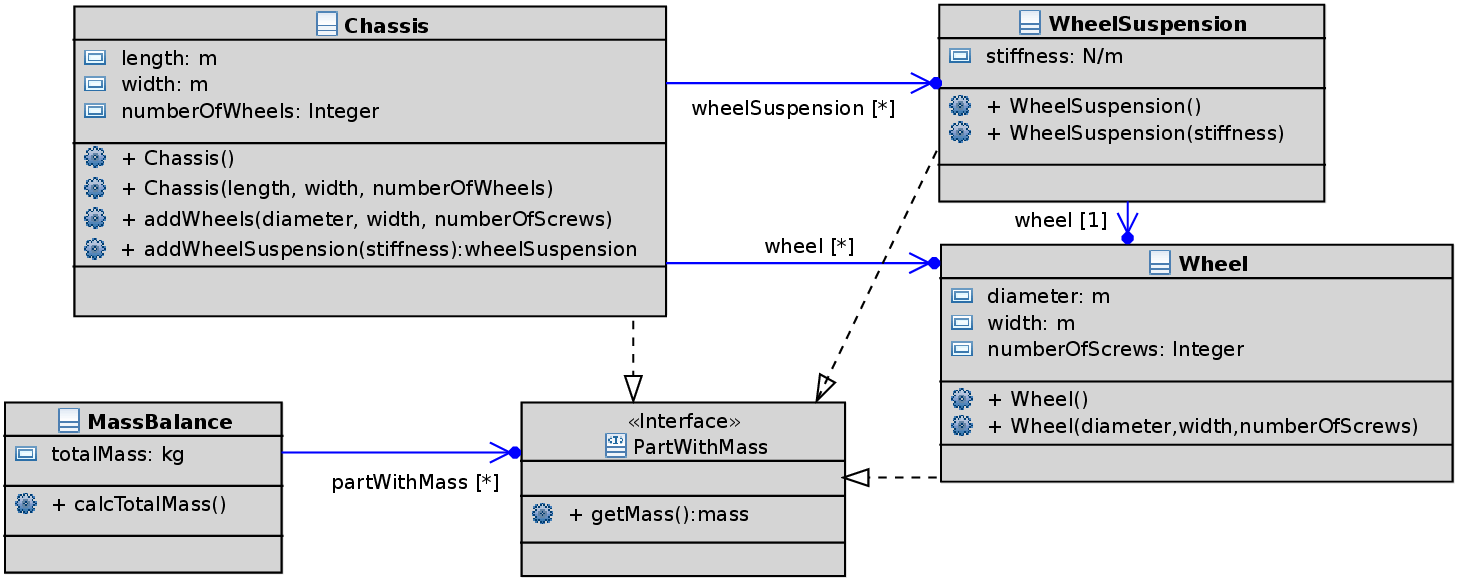}
\caption{The classes that realize an interface (dotted arrow) have to implement the methods defined in the interface.}
\label{fig:interface}
\end{figure}

An object-oriented interface mechanism is realized by the introduced class methods. Behavior of components is defined within interface elements by abstractly defined methods (empty methods with defined signatures). These interfaces are implemented by classes that realize the interfaces' abstract behavior. The interface mechanism provides \textit{abstraction} and \textit{polymorphism}. The behavior of components is modeled in an abstract way and sub-typing of entities through hierarchical inheritance and/or implementation relations is supported. Figure~\ref{fig:interface} shows an interface mechanism for calculating mass balances of mechanical parts. The parts inherit the behavior/property of having a mass from an interface. The method defined by the interface itself is used by a \textit{massBalance} class that calculates the total mass of the linked classes. 

\subsection{Calling Methods and Constructors in the Activity Diagram}
\label{subSec:callMethodsAndConstructors}
\begin{figure}[!ht]
\centering
\includegraphics[width=0.8\textwidth]{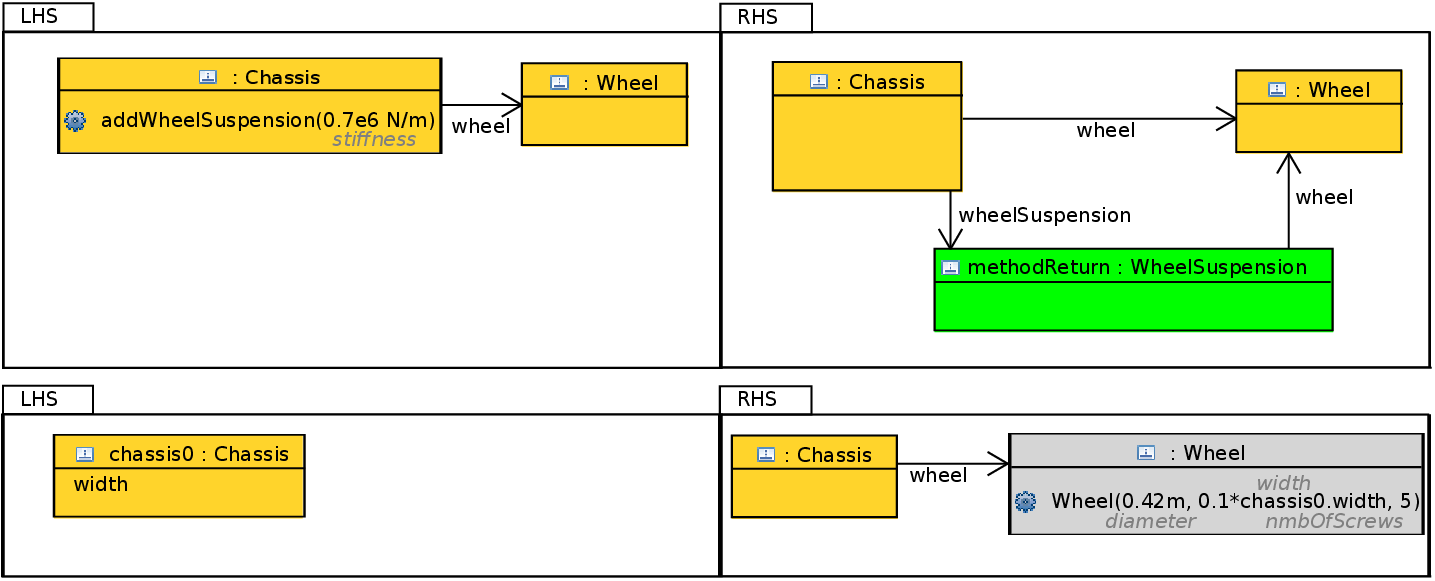}
\caption{Top: Calling method \emph{addWheelSuspension} that returns an instance \emph{methodReturn} which is integrated in the model on the RHS. Bottom: An instance of type \emph{Wheel} is created and linked to the existing \emph{Chassis} instance on the RHS by calling the constructor.}
\label{fig:callingMethodConstructor}
\end{figure}

According to subsection~\ref{subSec:codeIntegration} the design compiler provides graphical and code-based representations of the design grammar's components in parallel. Calling constructors and methods within Java code is self-explanatory, whereas a graphical way of calling constructors and methods needs to be defined to maintain the dual representations. Figure~\ref{fig:callingMethodConstructor} shows a mechanism to call a constructor (bottom) and a method (top) within graphical rules. The instance whose method shall be called, as well as the specific method, are defined on a graphical rule's LHS. Parameters in the method can be explicitly provided by value or referenced from other LHS instances' parameters. Possible return objects of the called method are created on the RHS of the graphical rule. Using this approach, the methods calls are used in the same spirit as conventional graphical rules. The context of a method call is defined on the LHS. The design graph change resulting from the call is defined on the RHS. Constructor calls are defined solely on the RHS, as a constructor creates an instance whose target context is searched on the LHS and the connection to the context is defined on the RHS. Parameters are defined as in the method call, either explicitly or via instances' parameters from within the rule context.

\subsection{Modeling Methods}
\label{subSec:modelingMethods}
\begin{figure}[!ht]
\centering
\includegraphics[width=0.8\textwidth]{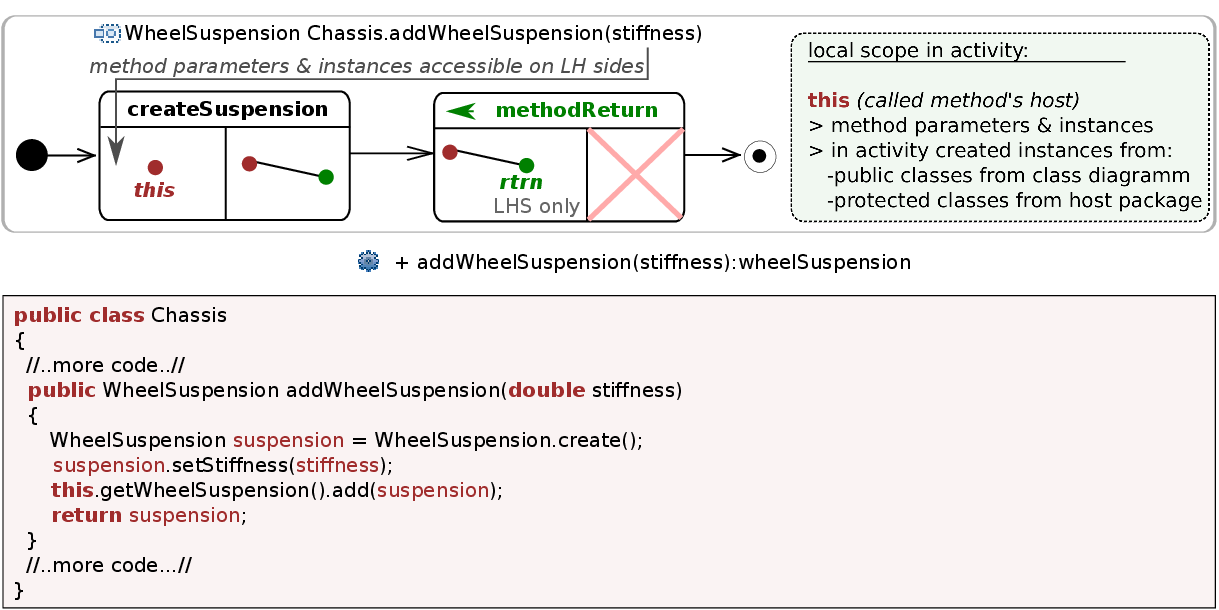}
\caption{Top: Graphical implementation of method \emph{addWheelSuspension} (method call in figure~\ref{fig:callingMethodConstructor}) as sub activity. The method's return instance or value is specified in a LHS search mechanism which is the sub activity's final element. Bottom: Method implemented as code rule within a specified method body.}
\label{fig:methodImplementation}
\end{figure}

The methods hull can now be filled either by Java code, as shown in figure~\ref{fig:methodImplementation} bottom, or graphically. The proposed graphical approach uses a sub activity and is shown in figure~\ref{fig:methodImplementation} top. The last element of the sub-activity is a fixed rule that is used for defining the method's return on the LHS - potentially within an LHS context. The methods parameters (instances or variables) are available within the sub activity's namespace. The variables or fields' values of the passed instance-type method parameters can be accessed by name. In the sub activity's rules the passed instances can be accessed and inserted explicitly via the context-assist of the rule editor. The accessibility of instances and variables in the method implementing sub activity complies with Java language, which accounts for \textit{encapsulation} and \textit{information hiding}.

\subsection{Encapsulating Design Langugages}
\label{subSec:subDesGram}
\begin{figure}[!ht]
\centering
\includegraphics[width=0.8\textwidth]{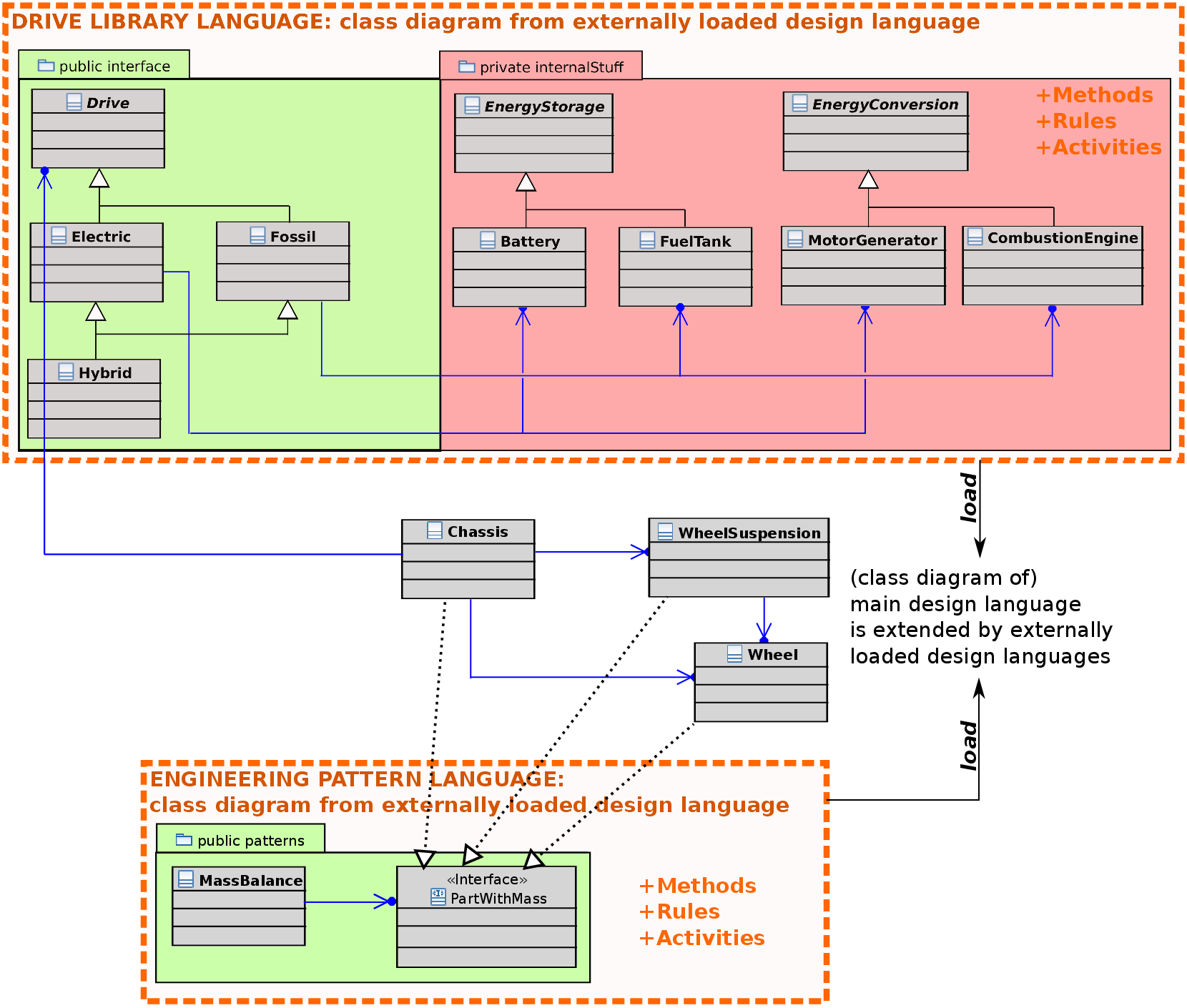}
\caption{Encapsulation and modularization: Extending a design grammar('s class diagram) through externally loaded design grammars (dotted frame) that are encapsulated modules. Only \emph{public} packages of the external design grammars are accessible by the main design grammar.}
\label{fig:multiClasses}
\end{figure}

Figure~\ref{fig:multiClasses} shows the proposed accessibility concept on the level of (re-)using and integrating multiple class diagrams from multiple design grammars. The classes in the class diagram are sorted to packages with different access modifiers. Only the classes and interfaces in the \textit{public} packages are shown and accessible from outer design grammars. This allows the creation of complex design tasks in hierarchically encapsulated sub design grammars. The internal implementation details are hidden from the outer design grammar that calls and uses the \textit{public} methods, classes and interfaces provided by the sub grammar.

%
%%%%%%%%%%%%%%%%%%%%%
\section{Results}
%\paragraph{What did you find?}
This section describes the findings in applying the presented methods from section~\ref{sec:methods}. Emphasis is put on addressing the issues from section~\ref{sec:problemSetting}.

\subsection{Information Security}
The presented methods allow an encapsulation and modularization of sub modules in sub design grammars as shown in section~\ref{subSec:subDesGram}. Together with the access modifiers from sections~\ref{subSec:modelingMethods} and~\ref{subSec:subDesGram}, this enables the encryption of (private) parts of sub design grammar modules with a distinct separation between the public (interface) elements, that are accessible by third parties, and the critical encrypted parts within the private packages (figure~\ref{fig:infoSecurityCollab}). Suppliers can share their design grammars without explicitly sharing their know-how and intellectual property.

\subsection{Collaboration}
\begin{figure}[!ht]
\centering
\includegraphics[width=0.8\textwidth]{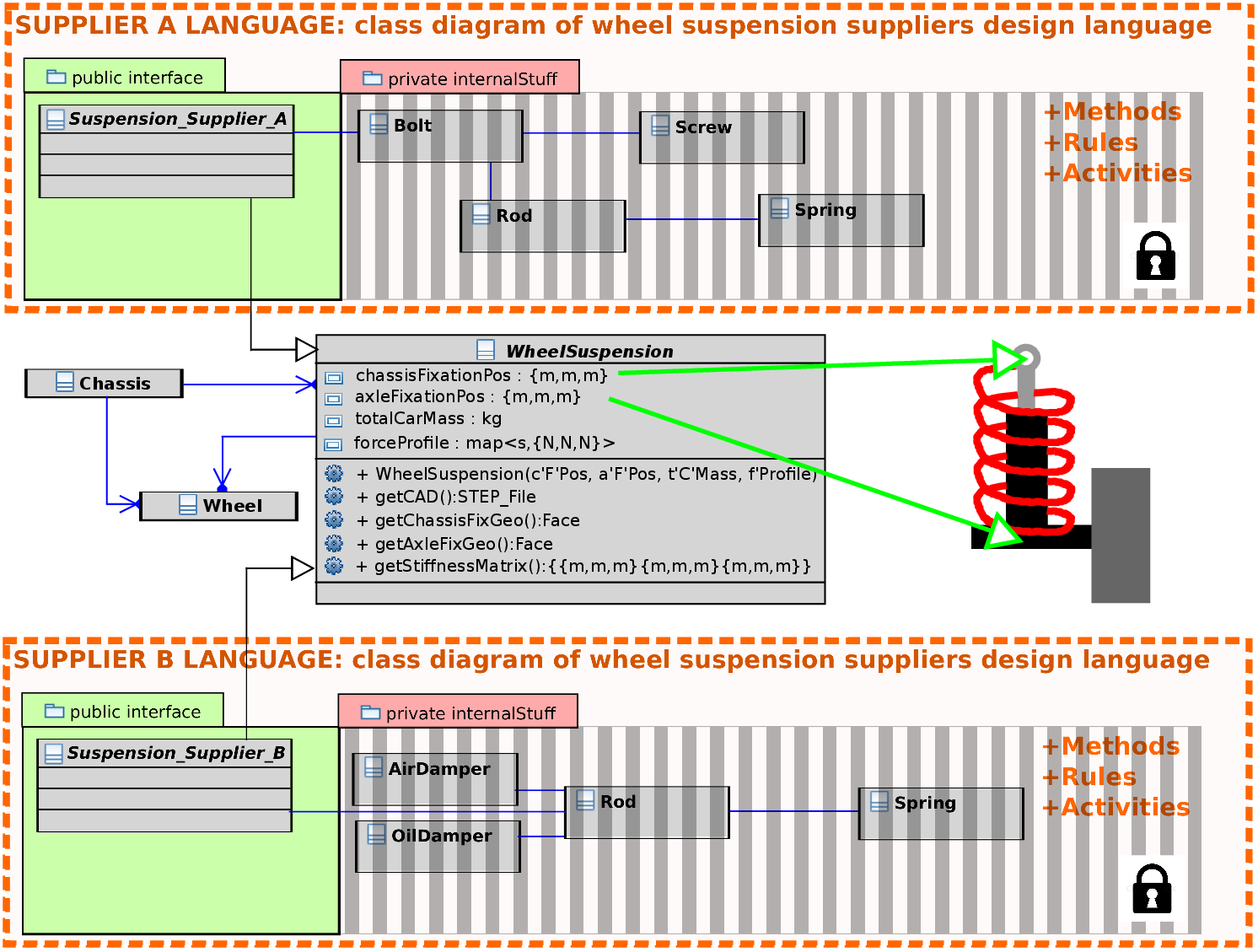}
\caption{Collaboration and division of labor by an embedding and composition of external design grammars. The public interfaces to the external design grammars are defined by interfaces or abstract classes. The activities, rules and classes in the private packages can be encrypted to ensure information security and know-how protection.}
\label{fig:infoSecurityCollab}
\end{figure}

A virtual blueprint of a product, consisting of OEM manufactured and third party components, can be realized this way. The integration of hierarchically organized sub design grammars (section ~\ref{subSec:subDesGram}) supports the collaborative creation of even complex design grammars in the cooperation between participants and experts from different domains and units (figure~\ref{fig:infoSecurityCollab}). The interface mechanism enables a distinct decomposition and assignment of tasks and subsystems between and to collaborators. The interfaces are explicitly modeled in the design grammar and are directly implemented by the responsible collaborators.

\subsection{Domain Integration}
\begin{figure}[!ht]
\centering
\includegraphics[width=0.8\textwidth]{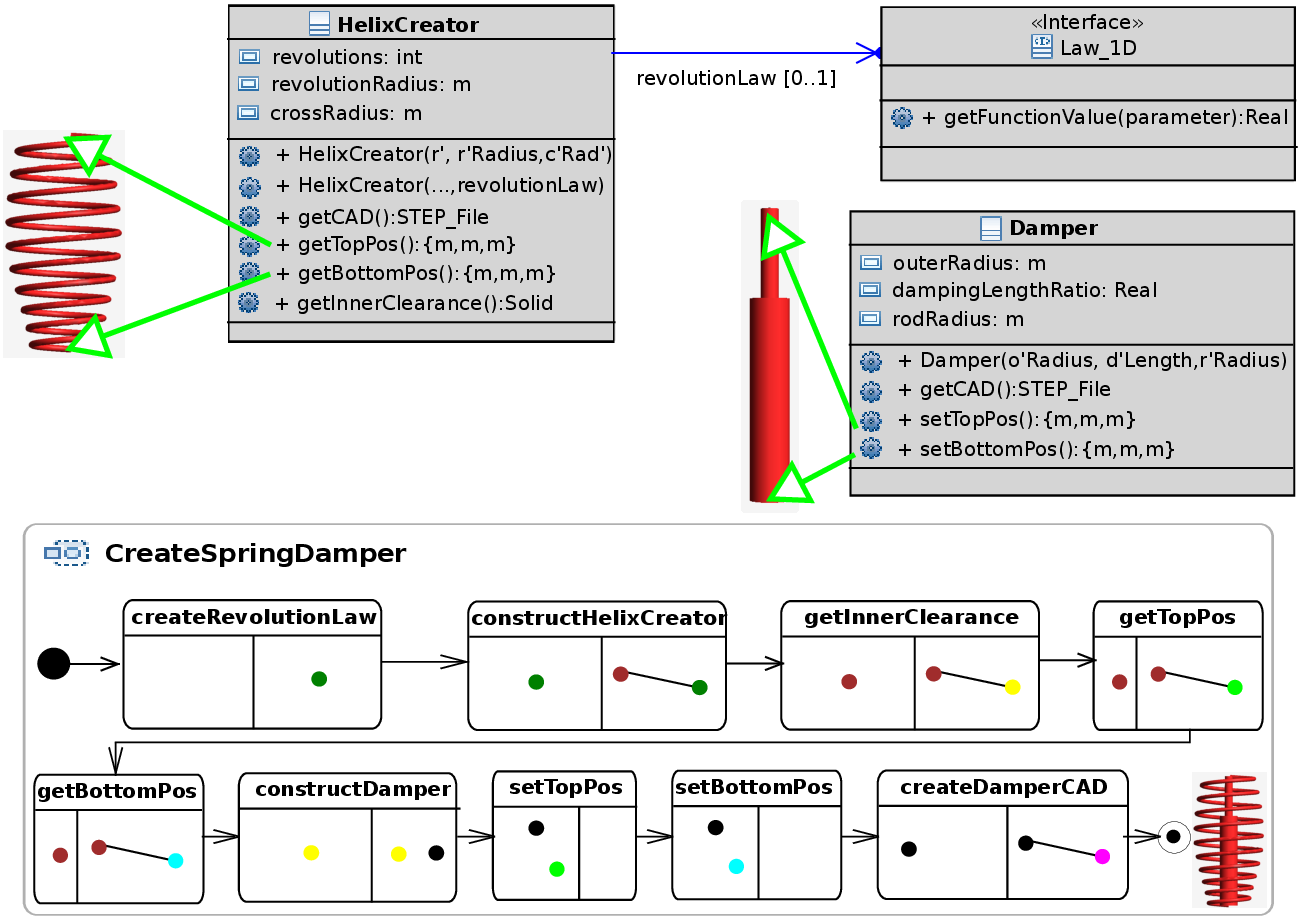}
\caption{Using the methods and interface mechanism to explicitly model an iterative CAD model creation (following operations use return values of preceding method calls as parameters). Top: class diagram; bottom: activity diagram with schematic graphical method and constructor calls(see figure~\ref{fig:callingMethodConstructor}).}
\label{fig:domainIntegration}
\end{figure}

The method call mechanism of section~\ref{subSec:callMethodsAndConstructors} supports the integration of different (physical) domains in the multidisciplinary product design. The CAD model creation is a central task in product design. In general, it is difficult to create a CAD geometry in one step. One step means, running a rule sequence that creates an abstract representation of the CAD geometry in the central model, which is translated to a CAD geometry within \textit{one CAD plug-in call} in the process chain. An iterative creation of the CAD model is much more suitable to create complex CAD models, as the subsequent geometry manipulations depend on the result of preceding geometry manipulations (figure~\ref{fig:domainIntegration}). For example are sketches extruded to bodies, whose faces are used as sketch planes for subsequent sketches and extrusions. This iterative modeling is generically realized with the presented class methods. They can be used to modularize and subtype behavior, as for example higher level CAD methods.

\subsection{Reusability}
\begin{figure}[!ht]
\centering
\includegraphics[width=0.8\textwidth]{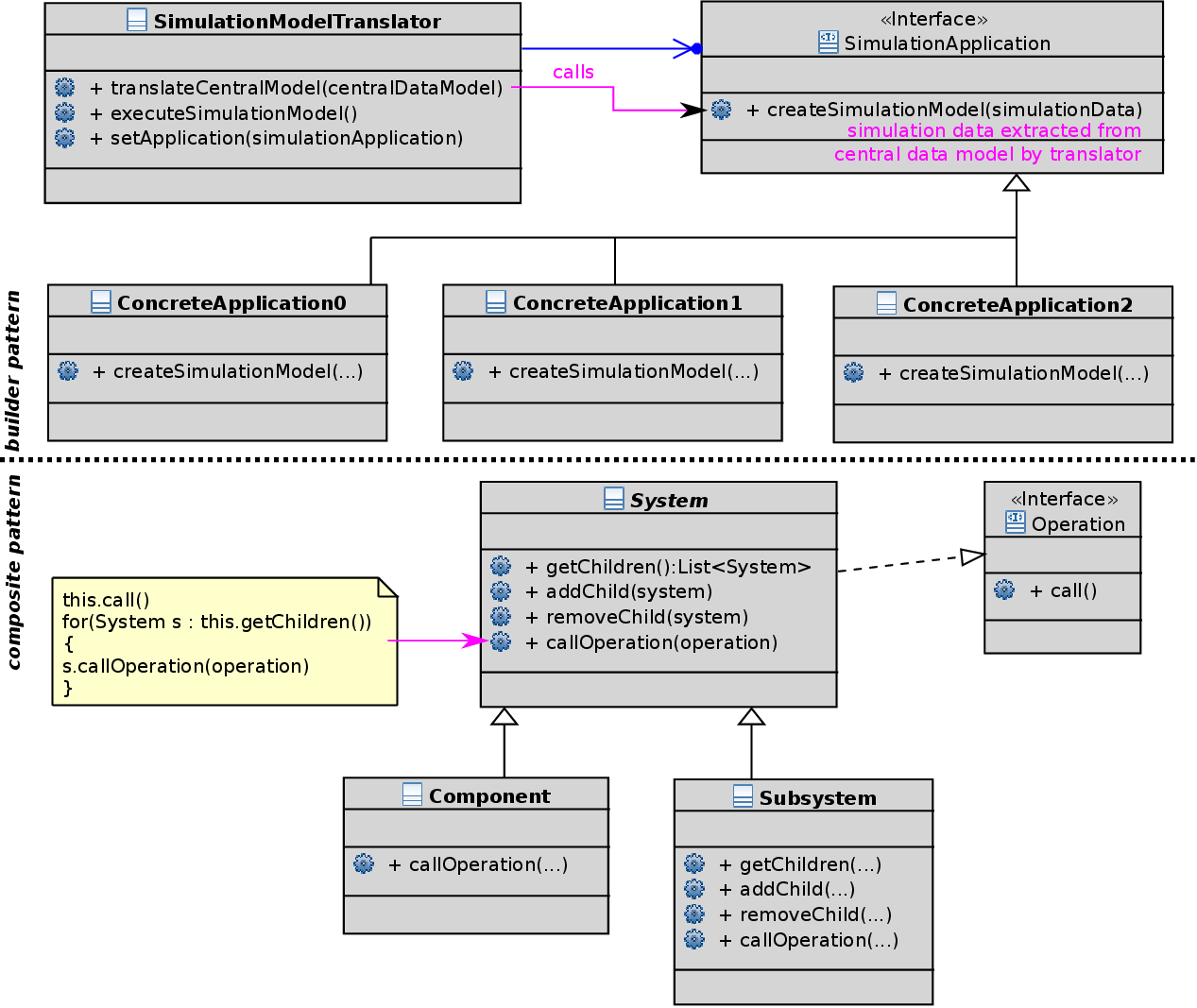}
\caption{Two design patterns from object-oriented software engineering. Top: \emph{Builder} pattern to translate the domain-relevant information in the central data model to different simulation applications (implementation of process chains in figure~\ref{fig:infoarchitectur}). Bottom: \emph{Composite} pattern to model systems-of-systems relationships in systems engineering.}
\label{fig:patterns}
\end{figure}

Design patterns are a well known strategy to create reusable software applications and modules~\cite{Gam1994} in the area of object-oriented software engineering. One can observe that all proposed patterns use \textit{interfaces} realizing the paradigm to \textit{''...program to an interface not an implementation...''}. Interfaces allow the definition of abstract and mandatory behaviors of components and modules that in turn can be re-used in many different contexts. Figure~\ref{fig:patterns} shows two applications of software engineering design patterns in the class diagram of a design grammar. A \textit{mass balance} engineering pattern is presented in figure~\ref{fig:multiClasses} to exemplary illustrate a potential application of the \textit{interface} mechanism to enable \textit{reusable product design patterns}.

\subsection{Modularization}
\begin{figure}[!ht]
\centering
\includegraphics[width=0.8\textwidth]{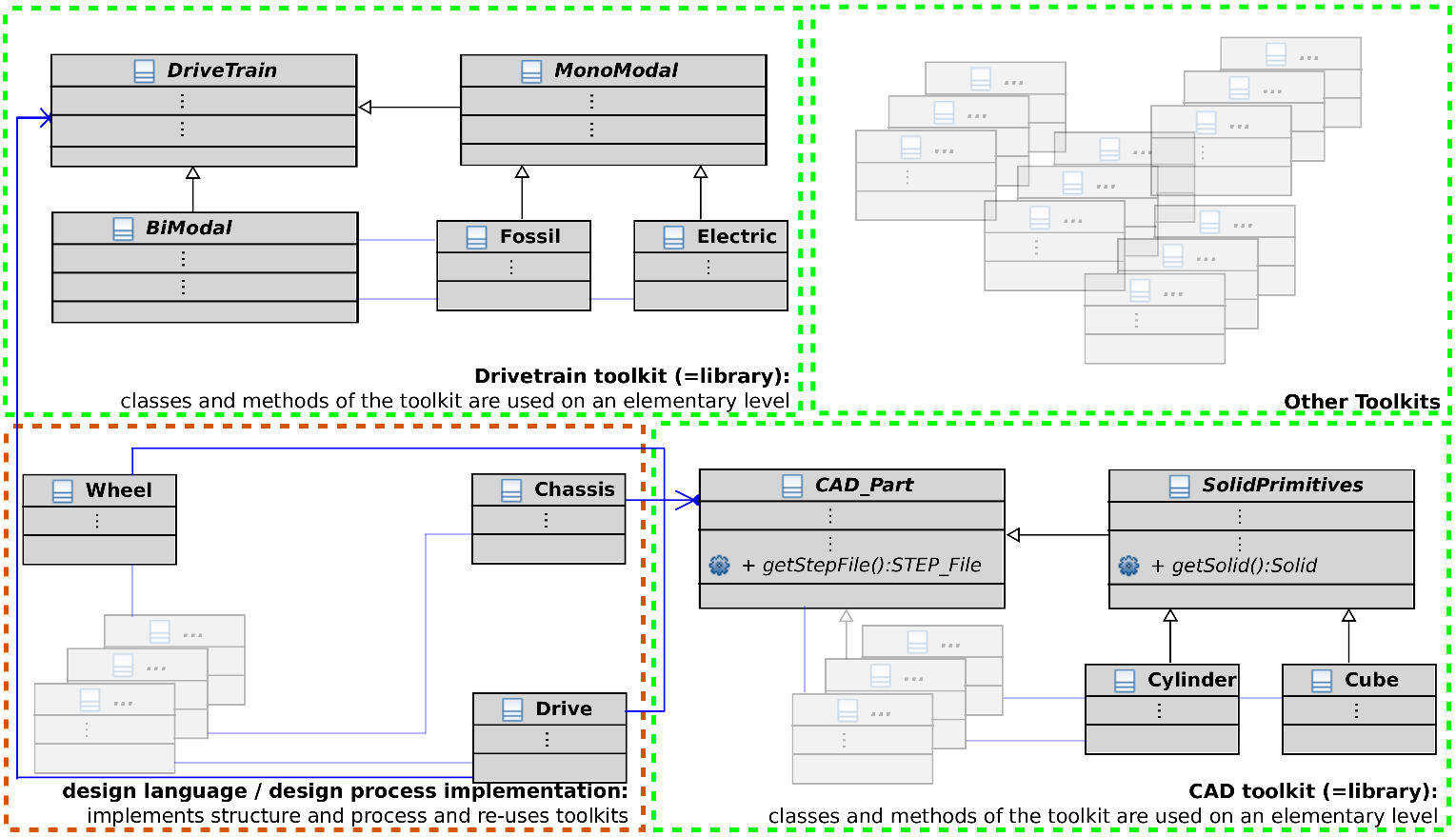}
\caption{Modeling the product structure and design process in a central design grammar. External toolkits for special purposes are embedded on an elementary level (single classes and methods) to re-use existing engineering solutions (toolkits=engineering libraries).}
\label{fig:toolkit}
\end{figure}

The introduction of interfaces and methods in the classes leads to an additional modularization on a higher level. Well known software design concepts as toolkits and software frameworks can now be realized in virtual engineering~\cite{Gam1994}. In a \textit{toolkit} setup, existing modules are reused as (software) libraries to achieve specific tasks. Transferring this approach to design grammars is exemplarily shown in figure~\ref{fig:toolkit}. 

\begin{figure}[!ht]
\centering
\includegraphics[width=0.8\textwidth]{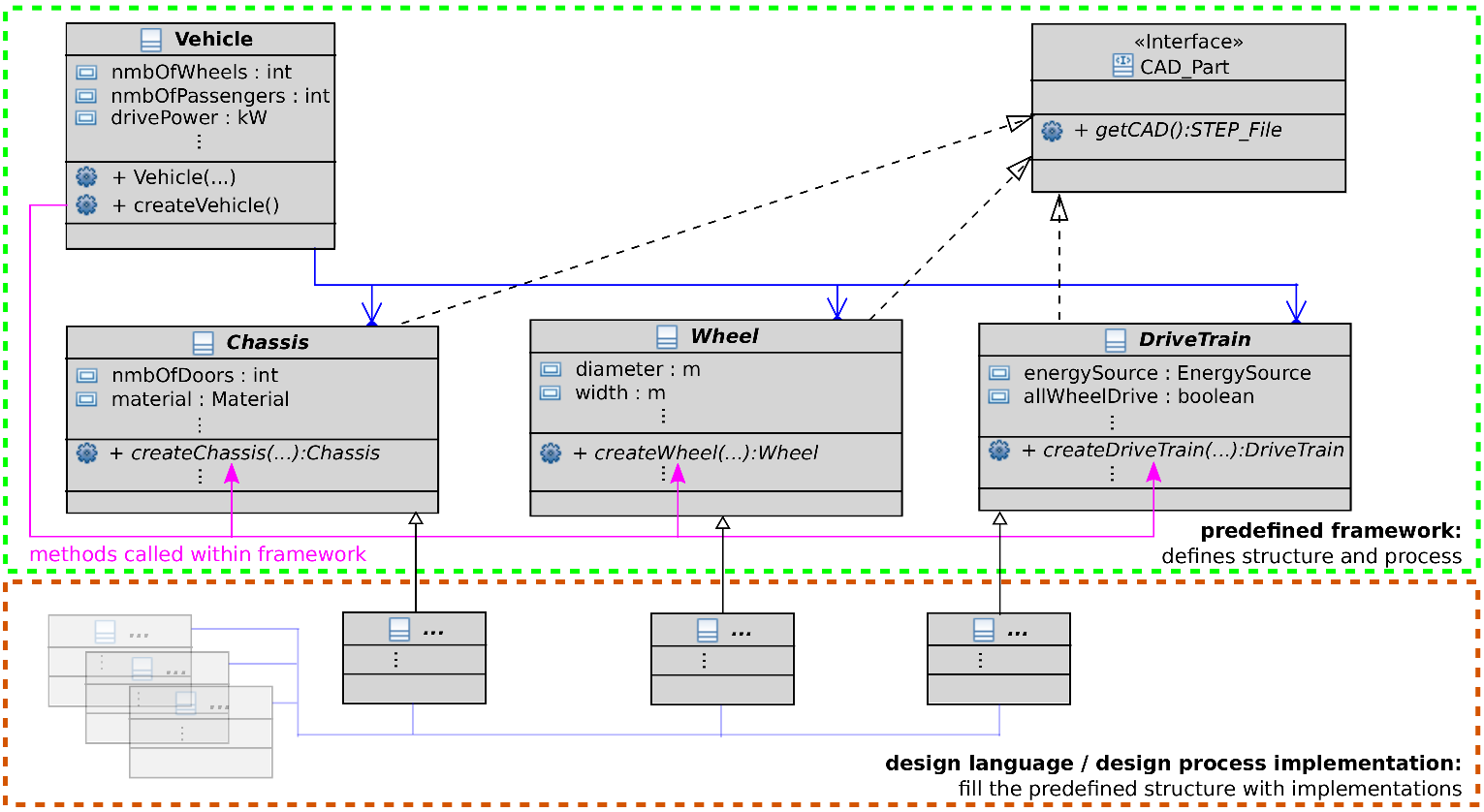}
\caption{A existing framework defines the product's structure and the product's design process through \emph{abstract interfaces} and \emph{abstract classes} (italic labels). The concrete design grammar implements the predefined methods and classes of the framework.}
\label{fig:framework}
\end{figure}

Figure~\ref{fig:framework} shows a \textit{framework}-based strategy of modularization. In the framework architecture, the global structure and decomposition is predefined by abstract \textit{interfaces} that have to be realized by the specific implementations. Transferred to product engineering, the design process is predefined (compare systems engineering~\cite{Inc2015}) in a framework which facilitates and structures the design process and enables exchange and reusability of existing components.

%%%%%%%%%%%%%%%%%%%%%
\section{Discussion}
%\paragraph{What does it all mean?}
Classical design grammars are put towards a higher level of object-orientation by introducing class methods and interfaces. A prototypically implementation within the Design Compiler 43 was created and graphical method execution mechanisms proposed, that fully support the dual programmatic and graphical modeling approach.

\paragraph{Conclusion}
The structuring and modularization of the product design process is profoundly improved and allows an much easier handling of the product's complexity in design and validation. The introduction of class methods and interfaces in design grammars enables the adoption of object-oriented design methods as design patterns and toolkit/framework architectures. Proven concepts as \textit{abstraction}, \textit{encapsulation} and \textit{polymorphism} are transfered to the virtual product design with design grammars. Design grammars are brought to the next level in terms of handling and automatizing complex product design processes.

\paragraph{Outlook}
The proposed mechanisms will be contained in the next release (DC43V3) of the Design Compiler. Based on properly modularized design processes, that are formally defined by their interface specifications, a self-organized design process should become feasible. This should work without an explicitly defined execution order in the central activity diagram. Starting from given product requirements, it seems possible to deduce an execution order of predefined modules in a self-organized manner, based on the module's interface signatures.\\

\textit{The work was partially supported by the project “digital product life-cycle (ZaFH)” funded by the European Regional Development Fund and the Ministry of
Science, Research and the Arts of Baden-Württemberg, Germany (www.rwb-efre.baden-wuerttemberg.de).}

%%%%%%%%%%%%%%%%%%%%%
\bibliography{./bibfile}

\end{document}